\documentstyle[12pt]{article}
\newlength{\aivwidth}   
\newlength{\tmpwidth}   

\newcommand{\phrd}[1]{Phys.\ Rev.\ {\bf D#1}}

\newcommand{\nphb}[1]{Nucl.\ Phys.\ {\bf B#1}}
\newcommand{\phlb}[1]{Phys.\ Lett.\ {\bf B#1}}
\newcommand{\zphc}[1]{Z.\ Phys.\ {\bf C#1}}

\newcommand{\eqnew}{\setcounter{equation}{0}}

\newcommand{\tfrac}[2]{\frac{\textstyle #1}{\textstyle #2}}
\newcommand{\suu}{$\rm SU(2)_L\times U(1)_Y$}
\newcommand{\nn}{\nonumber \\}
\newcommand{\be}{\begin{equation}}
\newcommand{\ee}{\end{equation}}
\newcommand{\bea}{\begin{eqnarray}}
\newcommand{\eea}{\end{eqnarray}}
\renewcommand{\O}[1]{{\cal L}_{#1}}
\newcommand{\tr}{\,{\rm tr}\,}
\newcommand{\lag}[1]{{\cal L}_{#1}}
\newcommand{\ep}[1]{\epsilon_{\scriptscriptstyle #1}}
\renewcommand{\d}{\partial}
\newcommand{\kg}{\kappa_\gamma}
\newcommand{\kz}{\kappa_Z}
\newcommand{\gz}{g_{ZWW}}
\newcommand{\lz}{\lambda_Z}
\renewcommand{\lg}{\lambda_\gamma}
\newcommand{\yg}{y_\gamma}
\newcommand{\yz}{y_Z}
\newcommand{\xg}{x_\gamma}
\newcommand{\xz}{x_Z}
\newcommand{\dz}{\delta_Z}
\newcommand{\kp}{\kg}
\newcommand{\hg}{\hat{g}}
\newcommand{\hhg}{\hat{\hat{g}}}
\newcommand{\tg}{\tilde{g}}
\newcommand{\sw}[1]{\sin^{#1} \theta_W}
\newcommand{\cw}[1]{\cos^{#1} \theta_W}
\newcommand{\egl}{{\textstyle\lambda}\frac{\textstyle e}{\textstyle g}}
\newcommand{\gel}{{\textstyle\lambda}\frac{\textstyle g}{\textstyle e}}
\newcommand{\eref}[1]{(\ref{#1})}

\newcommand{\zwe}{\left(\frac{M_Z ^2}{M_W ^2}-1\right)}
\newcommand{\sqzwe}{\sqrt{\frac{\textstyle M_Z^2}{\textstyle M_W^2}-1}}
\newcommand{\cotw}[1]{\frac{\cos^{#1}\theta_W}{\sin^{#1}\theta_W} }

\title{Non-Standard Gauge-Boson Self-Interactions\\within a Gauge Invariant
Model\thanks{Partially supported by Deutsche Forschungsgemeinschaft}}
\author{Carsten Grosse-Knetter\thanks{E-Mail:
knetter@physw.uni-bielefeld.de},\\
Ingolf Kuss\thanks{E-Mail: kuss@physf.uni-bielefeld.de}\\and\\Dieter
Schildknecht\\[5mm]Universit\"at Bielefeld\\Fakult\"at f\"ur Physik\\33501
Bielefeld\\Germany}
\date{BI-TP 93/15\\hep-ph/9304281\\April 1993\\(Updated Version)}

\begin{document}
\begin{titlepage}
\maketitle
\thispagestyle{empty}
\begin{abstract}
We examine dimension-six extensions of the standard electroweak Lagrangian
which are invariant under local \suu -transformations. The dimension-four
trilinear and quadrilinear effective interactions of the vector bosons with one
another are found to coincide with the vector boson interactions
previously derived from global SU(2) weak isospin symmetry broken by
electromagnetism. Supplementing the model by a well-known dimension-six
single-parameter quadrupole interaction leads to the most general
vector boson self-couplings that can be obtained by addition of dimension-six
terms to the standard Lagrangian. We examine in some detail another
\suu -symmetric interaction which contains $W_3 B$ mixing and modifies
both vector boson self-couplings and fermionic interactions.
Independently of being
strongly constrained by the LEP~1 data, the addition of this interaction
to the above-mentioned non-standard ones
does not change the form of the trilinear and quadrilinear non-standard
self-couplings
of the vector bosons. Therefore, while being
interesting in itself with respect to LEP~1 physics, this term is irrelevant
with respect to the phenomenology of the
vector-boson self-interactions.
\end{abstract}
\end{titlepage}


\section{Introduction}
\eqnew\typeout{Section 1}
The body of presently available data \cite{rol}
on electroweak phenomena is in excellent
agreement with the predictions of the standard \suu\ electroweak theory
\cite{sm}. The
agreement between theory and experiment is particularly noteworthy with respect
to the LEP~1 data which are precise enough to be sensitive to effects beyond
the simple tree approximation of the standard Lagrangian.

Concerning loop effects, it must be stressed, however, that the
agreement between theory and experiment is largely based on the
well-understood QED contribution of fermion loops to the photon propagator, the
``running'' of the electromagnetic fine structure constant, i.e., \
$\alpha(M_Z^2)=
\frac{1}{128.9}$, and the violation of global SU(2) symmetry induced by the
heavy-top-quark-loop. Quantitatively, this is seen by comparing
the exact one loop predictions with the ones of the
dominant-fermion-loop approximation \cite{gosch,kks}.
One finds that the contribution of the
$M_H$-dependent bosonic loop corrections is of the order of magnitude of the
LEP~1 experimental errors. The small bosonic loop corrections appear as a
consequence of the renormalizable non-Abelian \suu\
structure of the couplings of the vector bosons
with one another and the assumed existence of the Higgs scalar particle.
The consistency between theory and the LEP~1 precision data may
thus be considered to be a major triumph of the standard model, although the
bosonic loop corrections have not been quantitatively determined, as
evidenced by the uncertainty in the Higgs-boson mass, $M_H$, extracted
from the LEP~1 precision data.

The fact that the boson-loop vacuum-polarization effects
and, accordingly, the trilinear
and quadrilinear couplings among the vector bosons are not precisely
determined by
the LEP~1 data provides sufficient reason to perform direct measurements
of the trilinear self-couplings in $e^+e^-\to W^+W^-$ \cite{aa,bkrs} and of
the quadrilinear ones in $e^+ e^- \rightarrow W^+ W^- Z, W^+ W^-\gamma$
 \cite{gksch,bebo}.
Such a program can be started at LEP~2 and continued in the more distant future
at a linear $e^+e^-$ collider with an energy of, e.g.,  $\sqrt{s}=500\,$GeV
(NLC) \cite{bmt,base}.

On the other hand, even though any theory which correctly describes
the interactions of the vector bosons with the leptons and the quarks and which
yields sufficiently small
deviations from the standard model in the sector of gauge-boson
self-interactions would be compatible with the LEP~1 data, it is evidently
an important
question to be asked, whether at all we can think of alternatives to the
standard-model couplings among vector bosons which fulfil the LEP~1
constraints.

Some time ago, it was suggested (KMSS model \cite{kmss,schiaa})
that vector-boson self-couplings alternative to
the standard ones can indeed be introduced with a certain amount of
credibility, provided a minimum amount of symmetry is required for
the couplings in addition to Lorentz
invariance and electromagnetic gauge invariance. Restricting oneself to
dimension-four terms in the bosonic Lagrangian and requiring that global SU(2)
weak isospin symmetry be broken by electromagnetism only (by $\gamma-W^3$
current mixing, in particular),
one easily reconstructs the standard model in the
fermion sector, but obtains an alternative to the standard
model for the
trilinear couplings of the vector bosons to one another. While
Lorentz invariance and electromagnetic gauge invariance
in general allow for {\em three}
independent dimension-four trilinear couplings
among the vector bosons, the
underlying global SU(2) weak isospin
symmetry in the KMSS model restricts the number of free couplings
to {\em two} independent ones. Dimension-four trilinear and quadrilinear
self-couplings altogether are described in terms of {\em four} free
parameters. The standard \suu\ theory is recovered for a specific choice
of these four parameters.

In the present work, we will examine a
dimension-six
extension of the standard model, which is invariant under local
\suu\ transformations. It contains
two free coupling constants and the Higgs
scalar doublet in an essential manner. We will find that the trilinear
couplings of the vector bosons to one another, and also the quadrilinear
couplings of the abovementioned KMSS model, provided
certain qualifications are valid,
are precisely reproduced by this dimension-six
\suu\ invariant extension of the standard
Lagrangian. Originally \cite{kmss},
no use of the Higgs scalar was made, deliberately,
in the ansatz for the KMSS Lagrangian. One may in fact even challenge the
very idea of
introducing dimension-six nonrenormalizable Higgs interactions, as the
Higgs scalar was introduced originally to assure renormalizability.
Nevertheless, the introduction of non-standard Higgs interactions in the KMSS
model reduces the
cut-off-dependent bosonic loop divergences in physical quantities to
logarithmic ones
and, accordingly, protects its
predictions from disagreeing with the LEP~1 empirical
data \cite{haze,heve}.
On the other hand, in $e^+ e^- \rightarrow W^+ W^-$ and $e^+ e^-
\rightarrow W^+ W^- (\gamma ,Z^0 )$, the non-standard couplings lead to
deviations from the standard-model predictions which increase with the
square of the $e^+ e^-$ center-of-mass energy. The KMSS model thus
provides an excellent testing ground when empirically exploring the
vector-boson self interactions by measuring $e^+ e^- \rightarrow
W^+ W^-$, a reaction in which the Higgs interactions are negligible.
 It will allow one to precisely
quantify \cite{bkrs} to what extent future data agree with, e.g., the
standard-model predictions.

A more general model with non-standard couplings among the electroweak
vector bosons is obtained if the KMSS Lagrangian is extended by adding
a dimension-six quadrupole interaction among the vector bosons.
It is well known \cite{quad} that the appropriate use of non-Abelian
field-strength tensors allows one to introduce a quadrupole interaction
which is invariant under local \suu\ transformations and determined by a
single free parameter for both the $\gamma W^+ W^-$ and the $Z^0 W^+
W^-$ interactions. We will convince ourselves that the resulting
{\em three}-free-parameter model is indeed the most general
$SU(2)_L\times U(1)_Y$-symmetric
dimension-six
extension of the standard theory containing non-standard vector-boson
self-interactions, provided the vector-boson-fermion sector is required
to remain unmodified and higher derivatives as well as $C$- and $CP$-
violating terms are excluded.

Another single-parameter dimension-six extension of the standard model,
which, however,  affects vector-boson fermion interactions apart from the
interactions of the vector bosons among each other, has recently
received much attention \cite{ruj}. Essentially it consists of adding a
$W^3 B$ current-mixing term to the standard Lagrangian. The mixing term
is accompanied by non-standard Higgs interactions and non-standard
trilinear couplings of the vector-bosons to one another in such a manner that
the resulting interaction becomes invariant under local \suu\
transformations. We will demonstrate that the $W^3 B$ current-mixing
term in the Lagrangian can be removed by an appropriate diagonalization
procedure. Apart from the (standard and non-standard) interactions of
the Higgs scalar particle, the model, upon diagonalisation,
turns out to be a revival of the
Hung-Sakurai Lagrangian \cite{hs}. This model in itself is of much
interest, as it allows one to quantify \cite{kks,su2v} deviations from
the standard \suu\ symmetry of vector-boson fermion interactions.
For $e^+ e^- \rightarrow W^+ W^-$, the Hung-Sakurai model leads to a
fairly decent high-energy behaviour \cite{hs}, a feature rediscovered in
\cite{ruj} and erraneously uniquely associated with the local \suu\ symmetry
obtained when introducing non-standard Higgs interactions in addition to
formulating $W^3 B$ current mixing with the non-Abelian W-field tensor.
The additional Higgs interactions are irrelevant, however,
for the behaviour of the cross section for $e^+ e^- \rightarrow W^+
W^-$.  Quite apart from the fact that $W^3B$ mixing is
strongly constrained \cite{kks,ruj,su2v} by the
LEP~1 data, it will be shown that the omission of an
additional $W^3 B$ mixing term does not lead to any loss of
generality concerning the empirical implications of the aforementioned general
three-parameter model with respect to non-standard
vector-boson self-couplings, since the form of the relations among the
non-standard couplings
is independent of the presence of $\O{WB}$.

In section 2, we briefly review the parametrization of the trilinear
couplings of the vector bosons among one another. In section 3, we
rederive the KMSS model by extending the standard model by interaction
terms, which are invariant under local \suu\ transformations.
In section 4, we add the \suu -invariant quadrupole term to the
Lagrangian and comment on the generality of the so-defined
three-free-parameter Lagrangian.
Various additional terms can be disregarded, as they can simply be
removed by field and coupling-constant redefinitions.
The recent claim \cite{ruj} that certain non-standard terms are a priori
``unnatural'' is commented upon.
Section 5 is devoted to a thorough discussion of the $W^3 B$
current-mixing extension of the standard model. Final conclusions will
be drawn in section 6.


\section{Parametrization of Trilinear Vector-Boson Self-Couplings}
\eqnew\typeout{Section 2}
If we disregard the theoretically somewhat remote CP-violating couplings
\footnote{Compare, however, ref. \cite{cpviol} for an unambiguous
empirical test of $CP$ invariance via comparison of certain $W^+$ and
$W^-$ density-matrix elements.}
of the vector bosons with one another and a C-violating anapole term, the most
general trilinear couplings of electroweak vector bosons
are contained in the effective Lagrangian
(e.g., \cite{gg})
\bea \lag{3GB}&=&-ie[A_\mu(W^{-\mu\nu} W^+_\nu-W^{+\mu\nu}W^-_\nu)+
\kg A_{\mu\nu}W^{+\mu}W^{-\nu}]\nn
&&-ie\gz[Z_\mu(W^{-\mu\nu} W^+_\nu-W^{+\mu\nu}W^-_\nu)+
\kz Z_{\mu\nu}W^{+\mu}W^{-\nu}]\nn
&&+ie\frac{\lg}{M_W^2}A_\mu^{\,\,\,\nu} W_\nu^{-\lambda} W_\lambda^{+\mu}\nn
&&+ie
\frac{\cw{}}{\sw{}}\frac{\lz}{M_W^2}
Z_\mu^{\,\,\,\nu} W_\nu^{-\lambda} W_\lambda^{+\mu},\label{cubic}\eea
where $W^+_{\mu\nu}=\d_\mu W^+_\nu-\d_\nu W^+_\mu$, etc., and $\gz$, $\kg$ and
$\kz$ denote the $ZW^+W^-$ coupling and the electromagnetic and weak dipole
moment, respectively. The strengths of the dimension-six quadrupole terms, not
present in the standard model, are denoted by $\lg$ and $\lz$. It is convenient
to rewrite \eref{cubic} in a form \cite{bkrs} which is linear in the
deviations from the
standard couplings (given by $\gz=\frac{\sw{}}{\cw{}}$, $\kg=\kz=1$ and
$\lg=\lz=0$), i.e.,
\bea \lag{3GB}&=&-ie[A_\mu(W^{-\mu\nu} W^+_\nu-W^{+\mu\nu}W^-_\nu)+
A_{\mu\nu}W^{+\mu}W^{-\nu}]\nn&&
-ie\xg A_{\mu\nu}W^{+\mu}W^{-\nu}\nn&&
-ie(\frac{\cw{}}{\sw{}}+\dz)[Z_\mu(W^{-\mu\nu} W^+_\nu-W^{+\mu\nu}W^-_\nu)+
Z_{\mu\nu}W^{+\mu}W^{-\nu} ]\nn&&
-ie\xz Z_{\mu\nu}W^{+\mu}W^{-\nu}\nn&&
+ie\frac{\yg}{M_W^2}A_\mu^{\,\,\,\nu} W_\nu^{-\lambda} W_\lambda^{+\mu}\nn&&
+ie\frac{\yz}{M_W^2}
Z_\mu^{\,\,\,\nu} W_\nu^{-\lambda} W_\lambda^{+\mu}\label{ccubic}\eea
where
\bea
\dz&=&\gz-\frac{\cw{}}{\sw{}},\nn
\xg&=&\kg-1,\nn
\xz&=&\gz(\kz-1)\label{devis},\nn
\yg&=&\lg,\nn
\yz&=&\frac{\cw{}}{\sw{}}\lz.\label{para}\eea
In \eref{cubic} to \eref{para} $\sw{}\equiv\frac{e}{g}$, where $e$ is the
electromagnetic coupling strength and $g$ denotes the coupling of the
$W$-field to the weak isospin current.
Non-standard trilinear couplings will be expressed in the form \eref{ccubic}
in the subsequent sections.


\section{The KMSS Model}
\eqnew\typeout{Section 3}
Several years ago, a model of trilinear and quadrilinear vector-boson
self-interactions was suggested  \cite{kmss}
which contains four free parameters, two of
these in the trilinear sector.
The model is based on the minimal set of
assumptions necessary for an empirically succesful description of the
vector-boson--fermion
interactions. It extrapolates into the bosonic sector of the
theory. In the bosonic sector all dimension-four interactions among vector
bosons which fulfil a global SU(2) weak-isospin symmetry are allowed. This
symmetry is broken by the (primordial) photon,
in particular via mixing in the neutral
sector. For an appropriate choice of the mixing strength, one recovers the
standard vector-boson--fermion interactions, but obtains
non-standard interactions of
the vector bosons with one another. The latter
ones are given by \cite{kmss}
\bea \lag{KMSS}&=&
-ieA_\mu(W^{-\mu\nu} W^+_\nu-W^{+\mu\nu}W^-_\nu)-ie\kp
A_{\mu\nu}W^{+\mu}W^{-\nu}
\nn&&+i\left(e\frac{\sw{}}{\cw{}}-\frac{\hg}{\cw{}}\right)
Z_\mu(W^{-\mu\nu} W^+_\nu-W^{+\mu\nu}W^-_\nu)\nn&&
+i\left(e\kp\frac{\sw{}}{\cw{}}-\frac{\hg}{\cw{}}\right)
Z_{\mu\nu}W^{+\mu}W^{-\nu}
\nn&&
-e^2(A_\mu A^\mu W^+_\nu W^{-\nu} -A_\mu A_\nu W^{+\mu}
W^{-\nu})\nn&&
+2e\left(e\frac{\sw{}}{\cw{}}-\frac{\hg}{\cw{}}\right)
(A_\mu Z^\mu W^+_\nu W^{-\nu}-
\frac{1}{2}A_\mu Z_\nu (W^{+\mu} W^{-\nu}+W^{-\mu} W^{+\nu}))
\nn&&-\left[e\left(e\frac{\sw{2}}{\cw{2}}-2\hg\frac{\sw{}}{\cw{2}}\right)
+\frac{\hhg}{\cw{2}}\right]
(Z_\mu Z^\mu W^+_\nu W^{-\nu} -Z_\mu Z_\nu W^{+\mu} W^{-\nu})\nn&&
-\frac{\tg}{\cw{2}}Z_\mu Z^\mu W_\nu^+ W^{-\nu} -\frac{\tg}{4\cw{2}}
Z_\mu Z^\mu Z_\nu Z^\nu \nn&&
+\frac{1}{2}\hhg
(W^+_\mu W^-_\nu W^{+\mu}W^{-\nu}-W^+_\mu W^-_\nu W^{-\mu}W^{+\nu})
-\tg W^+_\mu W^-_\nu W^{-\mu}W^{+\nu}.
\label{kmss}\eea
This Lagrangian contains an arbitrary coupling constant, $\hg$, which
characterizes the strength of the SU(2)-invariant trilinear vector boson
self-interactions as well as an
arbitrary magnetic dipole moment of the $W^\pm$
boson, $\kp$, and two parameters, $\hhg$ and $\tg$, which appear in the
quadrilinear interaction terms only.
 The trilinear part of \eref{kmss} is easily seen to have the form
\eref{ccubic} with
\bea \dz&=&\frac{\sw{}\tfrac{\hg}{e}-1}{\cw{}\sw{}},\nn
\xg&=&\kp-1, \label{expr}\eea
while $\xz$ is related to $\xg$ via
\be \xz=-\xg\frac{\sw{}}{\cw{}}.\label{rel}\ee
The quadrupole interactions are absent in \eref{kmss}, i.e. $\yg=\yz=0$.
Relation \eref{rel} is a consequence of requiring global
SU(2) symmetry in the limit of
vanishing coupling of the $W^\pm$ to the (primordial, unmixed) photon.
Relation \eref{rel} can
only be violated at the expense of allowing for an extra isolated term
of the form
\be W^3_{\mu\nu}W^{+\mu}W^{-\nu}\label{iso}\ee
which violates SU(2) symmetry
intrinsically, i.e., independently of the presence
of the (primordial) photon field in the neutral sector. To summarize, the
trilinear couplings in the KMSS model
are characterized by two free parameters, $\dz$ and $\xg$,
and the relation \eref{rel} expressing $\xz$ in terms of $\xg$.

We turn to a reconstruction of the KMSS Lagrangian \eref{kmss} from a
gauge-invariant
extension of the standard model. In particular, we consider the
Lagrangian
\be \lag{eff}=\lag{SM}+\ep{W\Phi}\frac{g}{M_W^2}\O{W\Phi}
+\ep{B\Phi}\frac{g'}{M_W^2}\O{B\Phi},
\label{lageffkmss}\ee
containing the two $SU(2)_L\times U(1)_Y$-symmetric
dimension-six terms\footnote{Both interactions, \eref{owp} and \eref{obp},
are contained in
the systematic classification of non-standard interaction terms given in
\cite{buwyllr}.}
\be\O{W\Phi}=i\tr[(D_\mu\Phi)^\dagger W^{\mu\nu}(D_\nu\Phi)]\label{owp}\ee
and
\be\O{B\Phi}=-\frac{1}{2}i\tr[\tau_3
(D_\mu\Phi)^\dagger(D_\nu\Phi)] B^{\mu\nu}.\label{obp}\ee
Here, $\Phi$ denotes the standard complex scalar Higgs doublet field,
\be \Phi=\frac{1}{\sqrt{2}}((v+H){\bf 1}+i\varphi_i\tau_i).\ee
Its covariant derivative is given by
\be D_\mu \Phi=\partial_\mu\Phi+igW_\mu\Phi-\frac{i}{2}g'\Phi\tau_3B_\mu,\ee
where
\be W_\mu=\frac{1}{2}W_{\mu i}\tau_i,\ee
denotes the non-Abelian vector field, and
\bea W_{\mu\nu}&=&\partial_\mu W_\nu-\partial_\nu W_\mu+ig[W_\mu,W_\nu],\nn
B_{\mu\nu}&=&\partial_\mu B_\nu-\partial_\nu B_\mu\nu\label{not}\eea
are the field strength tensors.
As usual, $g'$ denotes the $\rm U(1)_Y$ coupling, $g'=\frac{e}{\cw{}}$.
Passing to the physical photon and $Z^0$ fields in \eref{lageffkmss},
one finds that indeed the
trilinear boson sector has the form \eref{ccubic} with
\bea\dz&=&\frac{\ep{W\Phi}}{\sw{}\cw{}},\nn
\xg&=&\ep{W\Phi}+\ep{B\Phi},\label{parameters1}\eea
and the constraint
\be \xz=-\xg\frac{\sw{}}{\cw{}},\label{rel2}\ee
which is identical with \eref{rel}.
The trilinear vector-boson interactions described by the
\suu -invariant Lagrangian \eref{lageffkmss} indeed
coincide with the trilinear vector-boson interactions of the KMSS
model, \eref{kmss}, based on global SU(2) symmetry broken by
electromagnetism. This result is of no surprise, as the
dimension-six interactions \eref{owp} and \eref{obp} upon
spontaneous symmetry breaking fulfil the requirement of restoration
of global SU(2) symmetry for vanishing hypercharge coupling
constant, $g^\prime$. Indeed, the contribution of $\O{B\phi}$
to the Lagrangian \eref{obp} vanishes in this limit, while the
trilinear couplings contained in $\O{W\phi}$ have the form
\be
\vec W^{\mu\nu} ( \vec W_\mu \times \vec W_\nu ),\label{vec}
\ee
i.e., there is no isolated term of form \eref{iso} present, which would
violate SU(2) symmetry intrinsically\footnote{Note that a term of the form
\eref{iso} can only be obtained by introducing an additional
dimension-eight interaction into the Lagrangian, which is given by
$i\tr[\tau_3 (D_\mu\Phi)^\dagger (D_\nu\Phi)]\tr[\Phi^\dagger
W_{\mu\nu}\Phi\tau_3 ]$ \cite{gore}. This term indeed yields an
extra term of the form \eref{iso}, which violates SU(2) symmetry
intrinsically.}.

The quadrilinear couplings contained in the Lagrangian \eref{lageffkmss}
(apart from terms of order $\dz^2$) coincide with the special case of the KMSS
model obtained by imposing the conditions
\bea
\hat{\hat g} &=&\hat g^2 , \nn
    \tilde{g}&=&0 \label{hat}
\eea
which reduce the number of four free parameters to {\em two} independent
ones, $\delta_Z$ and $x_\gamma$.

Using the connection \eref{parameters1} between the parameters
$\delta_Z , x_\gamma$ and $\ep{W\phi} , \ep{B\phi}$, the
Lagrangian \eref{lageffkmss} may be rewritten in a form which directly
allows one to read off its properties with respect to the
trilinear boson couplings,
\be
\O{} = \O{SM} + \frac{g}{M_W^2}\left(
\dz\sw{}\cw{}\O{W\phi} + (\xg\frac{\sw{}}{\cw{}}
-\dz\sw{2})\O{B\phi}\right) .\label{calli}\ee

Two special cases are worth noting. The omission of the
interaction of the hypercharge field \cite{gksch},
$\O{B\phi}$, corresponds to
\be
x_\gamma = \delta_Z \sin \theta_W \cos \theta_W
\label{theta}
\ee
and, upon using \eref{rel2},
\be \xz=-\dz\sw{2}\ee
and reduces the model to a single-parameter ($\dz$) model for trilinear
and quadrilinear interactions among the vector bosons. In this case,
trilinear and quadrilinear interactions of the vector bosons with
one another are such that \cite{bks} the most strongly
unitarity-violating terms of order $s^2$ in tree amplitudes for the
scattering of massive vector bosons are absent. Equivalently, quartic
divergences in one-loop corrections to the $\rho$-parameter
\cite{nss} (without any Higgs contribution) are absent
if \eref{theta} is imposed. It was originally this requirement of a
(moderately) decent high-energy and loop behaviour by which the condition
\eref{theta} was introduced (BKS model, \cite{bks,nss}) in the KMSS
model.

A second special case of \eref{calli} amounts to the omission
of $\O{W\phi}$ via
\be
\delta_Z = 0.  \nn
\ee
In this case $x_\gamma$ only and $\xz = - \xg (\sin\theta_W /
\cos \theta_W)$ remain as non-standard couplings.

With respect to the quadrilinear coupling strengths we noted that
the \suu -symmetric Lagrangian \eref{lageffkmss} yields a two-parameter
reduction of the quadrilinear couplings in \eref{kmss} which is
characterized by the constraints \eref{hat}. For completeness, we remark
that the more general four-parameter Lagrangian \eref{kmss} for
the couplings of the vector bosons among each other can also be
obtained within an \suu\ symmetric ansatz, at the expense, however,
of allowing for dimension-eight terms. One has to add the terms
\bea \frac{\ep{D\Phi,1}}{M_W^4}\O{D\Phi,1}&=&-\frac{\ep{D\Phi,1}}{4M_W^4}
\tr(D_\mu\Phi^\dagger D_\nu\Phi) \tr(D^\mu\Phi^\dagger
D^\nu\Phi)\nn
\frac{\ep{D\Phi,2}}{M_W^4}\O{D\Phi,2}&=&-\frac{\ep{D\Phi,2}}{4M_W^4}
\tr(D_\mu\Phi^\dagger D^\mu\Phi) \tr(D_\nu\Phi^\dagger
D^\nu\Phi)
\eea
to the Lagrangian \eref{lageffkmss} with the identification
\bea \ep{D\Phi,1}&=&2\hg^2-\hhg ,\nn \ep{D\Phi,2}&=&\tg-2\hg^2+\hhg .\eea

In summary, the trilinear and (special values of) the quadrilinear
couplings of the vector bosons of the KMSS model are recovered
from a Lagrangian which is invariant under local \suu\ transformations at
the expense of introducing dimension-six terms containing
non-standard interactions involving the Higgs particle.
The interactions of the Higgs scalar
are obviously irrelevant for the predictions of the KMSS model for
$W^+W^-$ production in $e^+e^-$ annihilation at LEP2- and higher
energies. They are of relevance, however, as far as vector boson loops
within the KMSS model are concerned. A
recent analysis \cite{haze,heve} of loop contributions
indeed has explicitly verified that only (mild) logarithmic dependences on the
(necessary) cutoff $\Lambda$ remain at the one-loop level\footnote{According
to a most recent
analysis in ref.\cite{haze2}, obtained after completion of the present
work, this point can even be strengthened. In
\cite{haze2} it is shown that the logarithmic divergences can indeed
be absorbed in the renormalization of (standard and non-standard)
couplings in the Lagrangian. Even though the relevant non-standard
couplings are strongly restricted by the LEP~1 results, practically
no conclusion can be drawn on the vector-boson self-couplings, unless
one imposes additional assumptions about certain bare couplings.},
apart from a quadratic dependence
on the Higgs-boson mass, $M^2_H$. By supplementing the
non-standard two-free-parameter vector-boson interactions of the
KMSS model with appropriately chosen non-standard Higgs
couplings, one indeed protects the one-loop predictions of the KMSS model from
disagreeing with the LEP1 precision data
(provided realistic choices of the Higgs mass, $M_H\le 1\,\rm TeV$, are
adopted).

In the dimension-six extension of the KMSS model we restricted
ourselves to the standard linear realization of the \suu\ symmetry,
spontaneously broken via the Higgs mechanism.
It seems appropriate to add a remark on
the alternative of realizing the \suu\
symmetry in a non-linear way \cite{apbe,bulo,gkko}.
Such a non-linear representation is obtained by
replacing the Higgs doublet $\Phi$ in \eref{owp} and \eref{obp} by
an expression which is
nonpolynomial in the
pseudo-Goldstone fields,  via
\be \Phi\equiv \frac{1}{\sqrt{2}}((v+H){\bf 1}+i\varphi_i\tau_i)\to U\equiv
\frac{v}{\sqrt{2}}\exp\left(\frac{i\varphi_i\tau_i}{v}\right).\label{lnl}\ee
The model obtained via this substitution in \eref{owp} and \eref{obp} is
symmetric under local \suu\ transformations without containing a
physical Higgs particle, and the KMSS model is now obtained by choosing
the unitary gauge.
Due to the absense of the Higgs scalar, this formulation of the model
leads to quadratic divergences at the one-loop level
in LEP~I observables. The aforementioned  quadratic dependence
\cite{haze,heve}
of the loop corrections on the Higgs boson mass, $M_H ^2$, is converted
into a $\Lambda ^2$ divergence, since the nonlinear model is the limit
of the linear model for $M_H \to\infty$.


\section{The General Three-Parameter Model}
\eqnew\typeout{Section 4}

\subsection{The Three-Parameter SU(2)${}_{\bf L}\bf
\times$U(1)${}_{\bf Y}$-Symmetric Model}
Supplementing the Lagrangian \eref{lageffkmss} by a well known quadrupole term
\cite{quad} constructed from the non-Abelian field-strength-tensor
$W_{\mu\nu}$ \eref{not},
\be \O{W}\,\,\,=\,\,\,-\frac{2}{3}i
\tr (W_\mu^{\:\nu}W_\nu^{\:\lambda}W_\lambda^{\:\mu}),
\label{ow}\ee
we obtain
\be \lag{eff}=\lag{SM}
+\ep{W\Phi}\frac{g}{M_W^2}\O{W\Phi}
+\ep{B\Phi}\frac{g'}{M_W^2}\O{B\Phi}
+\ep{W}\frac{g}{M_W^2}\O{W}.
\label{lageff}\ee
It turns out that $\ep{W}$ in \eref{lageff} is identical to $\yg$ in
\eref{ccubic},
\be\yg=\ep{W},\label{parameters2}\ee
while \cite{quad}
\be\yz=\yg\frac{\cw{}}{\sw{}}.\label{couprel}\ee

We note that the loop effects on LEP~1 observables due to the quadrupole
interaction \eref{ow} were recently examined in \cite{ruj}. As expected
from the underlying local \suu\ symmetry, an only mild logarithmic cut-off
dependence remains after renormalization. In this sense all three
non-standard interactions in
\eref{lageff} are equivalent to each other, even though,
on the other hand,
$\O{W\Phi}$ and
$\O{B\Phi}$ lead to a quadratic dependence of LEP~1-observables
on $M_H^2$, while (obviously) no $M_H$-dependence is present in the loop
corrections from $\O{W}$.

We will convince ourselves that the addition of the quadrupole interaction to
the KMSS model yields the most general\footnote{Any effective
Lagrangian with arbitrary self interactions can be embedded into an
\suu\ symmetric framework \cite{gore,bulo,gkko} provided terms of dimension
higher than six are allowed. The restriction to dimension-six terms
is motivated by simplicity.
Moreover, terms of higher dimension,
despite an underlying gauge symmetry, presumably imply more seriously
divergent loop contributions,
or rather higher powers of the Higgs mass in observables, and
are theoretically and phenomenologically disfavoured.} \suu -symmetric
dimension-six extension
of the standard
model as far as the interactions of the vector bosons with one another are
concerned, provided a few restrictive assumptions
are introduced to be specified
immediately. When claiming the generality of \eref{lageff}, we are restricting
ourselves to adding C- and CP-conserving interactions only. Interactions
containing higher derivatives are omitted, and we have obviously
restricted ourselves to deviations from the standard model which leave the
empirically well-known fermion interactions untouched.

The omission of a few particular terms in \eref{lageff}, nevertheless,
needs to be
discussed when claiming generality of \eref{lageff}. First of all, there are
two dimension-six terms
\bea\O{WW,1}&=&-\frac{1}{4}
\tr(\Phi^\dagger \Phi)\tr(W_{\mu\nu}W^{\mu\nu}),\label{oww1}\\
\O{WW,2}&=&-\frac{1}{2}
\tr(\Phi^\dagger W_{\mu\nu}W^{\mu\nu}\Phi)\label{oww2},\eea
which contain the non-Abelian field-strength tensor $W_{\mu\nu}$. By carrying
out a suitable redefinition of the $W$ field and of the coupling $g$, the terms
\eref{oww1} and \eref{oww2} can be shown to be reduced to non-standard
interactions containing the Higgs scalar, while neither the fermionic
interactions nor the interactions of the vector bosons among each other are
affected. The terms \eref{oww1} and \eref{oww2} can be neglected without loss
of generality.

Another interesting dimension-six interaction is given by
\be
\O{WB}\,\,\,=\,\,\,-\frac{1}{4}
\tr(\Phi^\dagger W_{\mu\nu}\Phi \tau_3)B^{\mu\nu}
\label{owb}.\ee
It contains $W^3 B$ mixing and in addition a non-standard $B\,W^+ W^-$
interaction (apart from non-standard Higgs terms).
As mentioned in the introduction, this interaction, when added to the standard
Lagrangian, yields the Hung-Sakurai electroweak model \cite{hs}
supplemented by non-standard Higgs interactions.
The term \eref{owb}  is of interest with respect to LEP~1 physics
(in fact, its strength is strongly suppressed \cite{kks,su2v} from LEP~1
data), but
irrelevant with respect to non-standard boson couplings, as its addition
to \eref{lageff}
leaves the form of the non-standard vector-boson
self-interactions unchanged.
This will be discussed thoroughly in section 5.

With respect to the point of view recently expressed,
that the interactions $\O{W\Phi}$, $\O{
B\Phi}$ and $\O{W}$ are ``contrived'' \cite{heve} and ``unnatural''
\cite{ruj}, a few additional comments may
be appropriate. As an example, we consider $\O{W\Phi}$ as defined by
\eref{owp}. Identical
arguments can be given for $\O{B\Phi}$ and $\O{W}$.
The physical content of the theory described by the Lagrangian
\be \O{}=\O{SM}+\ep{W\Phi}\frac{g}{M^2_W}\O{W\Phi} \label{1}\ee
remains invariant if the $W^\mu$ field is replaced by
\be W_i^\mu\to W_i^\mu-\frac{i}{2}\ep{W\Phi}\frac{g}{M_W^2}
\tr[\Phi^\dagger\tau_i(D^\mu\Phi)]. \label{pt}\ee
The invariance of the theory
is due to the fact that a field transformation in the
path integral does not change the theory\footnote{This holds apart
from $\delta^4(0)$-terms, which arise from the Jacobian of the
transformation.}.
To order $\ep{W\Phi}$, the substitution \eref{pt}
yields the Lagrangian
\bea \O{}&=& \O{SM}+\ep{W\Phi}\frac{g}{M_W^2}\Big\{
g'\O{WB}-g\O{WW,2}\nn&&-\frac{1}{8}g[\tr(\Phi^\dagger\tau_i
(D_\mu\Phi))][\tr((D^\mu\Phi^\dagger)\tau_i\Phi)-\tr(\Phi^\dagger\tau_i(
D^\mu\Phi))+2i\bar{\Psi}_L\gamma^\mu\tau_i\Psi_L)]\Big\}\nn&&
+O(\ep{W\Phi}^2)+\delta^4(0)\mbox{ terms}\label{2}\eea
which is equivalent to \eref{1}, even though it looks rather complicated.
Formula \eref{2} and similar ones for $\O{B\Phi}$ and $\O{W}$ were given in
\cite{ruj}\footnote{The fact that \eref{2} is obtained by inserting
the equations of motion \cite{ruj} into the Lagrangian does not provide
a sufficient justification for the validity of \eref{2}.
Since the equations of motion must not be
substituted into the Lagrangian, one has, instead, to perform the
field transformation \eref{pt} to obtain the result \eref{2} \cite{eom}.}.

While $\O{W\Phi}$, according to its definition \eref{owp},
only yields bosonic loop effects on LEP~1 observables,
the form \eref{2} of the Lagrangian now suddenly contains terms which,
taken by themselves, contribute at tree level to LEP~1 observables,
since they involve non-standard fermion interactions.
One may explicitly convince oneself that the tree-level effects of
\eref{2} on LEP~1 observables
cancel, as obviously implied by the equivalence of
\eref{1} and \eref{2}. The fact that the interactions $\O{WB}$ and the
fermionic term in \eref{2}, when individually analyzed, are strongly
suppressed by the LEP~1 empirical results
thus appears to be irrelevant for the strength and possible presence of
$\O{W\Phi}$. In other words,
from the fact that certain fermion--boson terms are absent
one can hardly deduce that also purely non-standard bosonic
interactions are absent in nature. Indeed, it is not a priori excluded that
precisely the purely bosonic sector in which the non-Abelian structure
of the gauge group and
the spontaneous symmetry breaking via the Higgs mechanism are manifest
deviates from
orthodoxy. Direct empirical tests of the boson sector seem unavoidable.
Moreover, we note that an interaction term, such as $\O{W\Phi}$ \eref{owp},
can hardly be considered as ``unnatural'' \cite{ruj}, simply because,
by a fairly involved field transformation, it can be
cast into a lengthy and awkward form, such as \eref{2}.


\subsection{Non-Standard Gauge-Boson Interactions}
In this section, we briefly collect the various standard and non-standard
interactions implied by the three-parameter Lagrangian \eref{lageff}.
We will use the parameters $\dz$, $\xg$ and $\yz$, which are related to the
couplings $\ep{W\Phi}$, $\ep{B\Phi}$ and $\ep{W}$ by \eref{parameters1} and
\eref{parameters2}. We will restrict ourselves to the relevant trilinear and
quadrilinear boson couplings (in the unitary gauge, $\varphi_a\equiv 0$).
Multi-particle vertices of fairly remote
significance are not listed.

\subsubsection{Trilinear Gauge-Boson Self-Interactions\label{cubics}}
The trilinear couplings are parametrized as in \eref{ccubic}. With
\eref{rel}
and \eref{couprel} one finds
in terms of $\dz, \xg$ and $\yg$
\bea \lag{3GB}(\dz,\xg,\yg)&=&
-ie[A_\mu(W^{-\mu\nu} W^+_\nu-W^{+\mu\nu}W^-_\nu)+
A_{\mu\nu}W^{+\mu}W^{-\nu}]\nn&&
-ie(\frac{\cw{}}{\sw{}}+\dz)[Z_\mu(W^{-\mu\nu} W^+_\nu-W^{+\mu\nu}W^-_\nu)+
Z_{\mu\nu}W^{+\mu}W^{-\nu} ]\nn&&
-ie\xg \left(A_{\mu\nu}W^{+\mu}W^{-\nu}
-\frac{\sw{}}{\cw{}}Z_{\mu\nu}W^{+\mu}W^{-\nu}\right)\nn
&&+ie\frac{\yg}{M_W^2}\left(A_\mu^{\,\,\,\nu} W_\nu^{-\lambda} W_\lambda^{+\mu}
+\frac{\cw{}}{\sw{}}
Z_\mu^{\,\,\,\nu} W_\nu^{-\lambda} W_\lambda^{+\mu}\right).\eea
The three-gauge-boson interaction
differs from the standard model in the overall normalization of
the $Z^0$ coupling and in the non-standard dipole and quadrupole
terms, in which the relative strengths of the $Z^0$ and $\gamma$
couplings are fixed, however.

\subsubsection{Quadrilinear Gauge-Boson Self-Interactions}
$\O{B\Phi}$
does not contain quadrilinear gauge-boson self-interactions.
Thus, the non-standard quadrilinear couplings only depend
on $\ep{W\Phi}$ and $\ep{W}$ or,
alternatively, by use of \eref{parameters1} and
\eref{parameters2}, on $\dz$
and $\yg$.
$\O{W\Phi}$ simply changes the coupling constants of the Yang--Mills
interactions, while $\O{W}$ yields extra couplings containing field strength
tensors,
\bea\lag{4GB}(\dz,\yg)&=&
-e^2(A_\mu A^\mu W^+_\nu W^{-\nu} -A_\mu A_\nu W^{+\mu}
W^{-\nu})\nn&&
-2e^2\frac{\cw{ }}{\sw{ }}\left(1+\dz\frac{\sw{}}{\cw{}}\right)\nn&&\qquad
(A_\mu Z^\mu W^+_\nu W^{-\nu}-
\frac{1}{2}A_\mu Z_\nu (W^{+\mu} W^{-\nu}+W^{-\mu} W^{+\nu}) )
\nn&&-e^2\frac{\cw{2}}{\sw{2}}\left(1+2\dz\frac{\sw{}}{\cw{}}\right)
(Z_\mu Z^\mu W^+_\nu W^{-\nu} -Z_\mu Z_\nu W^{+\mu} W^{-\nu})\nn&&
+e^2\frac{1}{2\sw{2}}\left(1+2\dz\sw{}\cw{}\right)\nn&&\qquad\qquad
(W^+_\mu W^-_\nu W^{+\mu}W^{-\nu}-W^+_\mu W^-_\nu W^{-\mu}W^{+\nu})\nn&&
+e^2\frac{\yg}{M_W^2}\bigg\{
[A_\mu A^{\mu\lambda}(W^{+\nu}W^-_{\nu\lambda}+
W^{-\nu}W^+_{\nu\lambda})\nn&&\qquad\qquad
+A_\mu A_{\nu\lambda}
(W^{+\nu}W^{-\lambda\mu}+
W^{-\nu}W^{+\lambda\mu})]\nn&&\qquad
+\frac{\cw{ }}{\sw{ }}
[(A_\mu Z^{\mu\lambda}+Z_\mu A^{\mu\lambda})(W^{+\nu}W^-_{\nu\lambda}+
W^{-\nu}W^+_{\nu\lambda})\nn&&\qquad\qquad
+(A_\mu Z_{\nu\lambda}+Z_\mu A_{\nu\lambda})
(W^{+\nu}W^{-\lambda\mu}+
W^{-\nu}W^{+\lambda\mu})]\nn&&\qquad
+\frac{\cw{2}}{\sw{2}}
[Z_\mu Z^{\mu\lambda}(W^{+\nu}W^-_{\nu\lambda}+
W^{-\nu}W^+_{\nu\lambda})\nn&&\qquad\qquad
+Z_\mu Z_{\nu\lambda}
(W^{+\nu}W^{-\lambda\mu}+
W^{-\nu}W^{+\lambda\mu})]\nn&&\qquad
+\frac{1}{\sw{2}}(W^+_\mu W^{+\mu\lambda}W^{-\nu}
W^-_{\nu\lambda}-W^+_\mu W^{-\mu\lambda}W^{-\nu}
W^+_{\nu\lambda})\bigg\}.\label{quartic}\eea

\subsubsection{Interactions of two Gauge Bosons and one or two Higgs Scalars}
$\O{W}$ does not contain
Higgs terms, i.e.\ the non-standard
interactions involving the Higgs particle depend on
$\dz$ and $\xg$ only.
The interaction terms $\O{W\Phi}$ and $\O{B\Phi}$ extend the standard
couplings of
$WWH(H)$ and $ZZH(H)$ by couplings involving one field-strength tensor and a
derivative of the Higgs field, and they give rise to new
$Z\gamma H(H)$ vertices in addition,
\bea\lag{2GB,1H}(\dz,\xg)&=&e\frac{1}{\sw{ }}M_W
W_\mu^+ W^{-\mu}H+e\frac{1}{2\sw{ }\cw{2}}
M_WZ_\mu Z^\mu H\nn&&
+e\frac{1}{M_W}\dz\cw{}(W^+_{\mu\nu}W^{-\mu}+W^-_{\mu\nu}
W^{+\mu})\d^\nu H\nn&&
+e\frac{1}{M_W}\left(2\dz\sw{}-\xg\frac{1}{\cw{}}\right)
A_{\mu\nu}Z^\mu\d^\nu H\nn&&
+e\frac{1}{M_W}\left(\dz\left(\cw{}-\frac{\sw{2}}{\cw{}}\right)+\xg\frac{\sw{}}
{\cw{2}}\right)
Z_{\mu\nu} Z^\mu\d^\nu H,\label{ggh}\eea
\bea\lag{2GB,2H}&=&e^2\frac{1}{4\sw{2}}M_WW_\mu^+ W^{-\mu}H^2
+e^2\frac{1}{8\sw{2}\cw{2}}
M_WZ_\mu Z^\mu H^2\nn&&
+e\frac{1}{2 M_W}\dz\frac{\cw{}}{\sw{}}(W^+_{\mu\nu}W^{-\mu}+W^-_{\mu\nu}
W^{+\mu})(\d^\nu H)H\nn&&
+e^2\frac{1}{2 M_W^2}\left(2\dz-\xg\frac{1}{\sw{}\cw{}}\right)
A_{\mu\nu}Z^\mu(\d^\nu H) H\nn&&
+e^2\frac{1}{2 M_W^2}\left(
\dz\left(\frac{\cw{}}{\sw{}}-\frac{\sw{}}{\cw{}}\right)+\xg\frac{1}
{\cw{2}}\right)
Z_{\mu\nu} Z^\mu(\d^\nu H) H.\label{gghh}\eea


\subsection{Leading Amplitudes for $\bf e^+e^-\rightarrow W^+
W^-$ \label{lead}}
The effective Lagrangian \eref{lageff} spoils the decent (${O}(s^0)$)
high-energy behaviour of the tree-level helicity amplitudes for
$e^+ e^- \rightarrow W^+ W^-$ as given in the standard model.
The helicity amplitudes
for arbitrary parameters $\dz ,\xg ,\xz ,\yg$ and $\yz$
as defined in \eref{ccubic} may be found in \cite{bkrs} and
\cite{bmt}. Using the constraints \eref{rel},
 \eref{couprel}, we list the tree amplitudes
to ${O}(\sqrt{s})$ and higher.

For equal transverse polarizations of the $W$'s and lefthanded electrons
the leading amplitude
\be {\cal M}_{++}={\cal M}_{--}=\frac{g^2}{8}\sin\vartheta
\frac{s}{M_W^2}\yg\label{mtt}\ee
depends on the quadrupole term, only. For righthanded electrons the
amplitude is
${O}(s^0)$.The amplitude for two $W$'s with
opposite transverse polarizations is also ${O}(s^0)$.
The amplitude for two longitudinally
polarized $W$'s,
\bea{\cal M}_{LL}^{\sigma}&=&-\frac{e^2}{2}\sigma\sin\vartheta\frac{s}{M_W^2}
\left[\xg+\left( \frac{\sw{}}{\cw{}}\xg -\dz\right) (a-2b\sigma )\right]
,\label{mll}\eea
gets no contribution from the quadrupole term and is
determined by the modifications of the Yang-Mills couplings,
$\dz$ and $\xg$. It has the same high energy behaviour as the
amplitude \eref{mtt} for equal transverse polarizations.
For one longitudinally and one transversely polarized $W$ the amplitude
\be {\cal M}_{LT}=-\frac{e^2}{2\sqrt{2}}\sigma (\tau\cos\vartheta -2
\sigma )\frac{\sqrt{s}}{M_W}\left[ \xg +\yg +\left(\xg\frac{\sw{}}{\cw{}}
-2\dz -\yg\frac{\cw{}}{\sw{}}\right)(a-2b\sigma)\right],\ee
\bea {\rm with}\quad\tau&=&+1\quad\mbox{for $W^+(+)$ or $W^-(-)$}, \nn
 \tau&=&-1\quad\mbox{for $W^+(-)$ or $W^-(+)$}, \eea
(where the signs in parentheses after the $W^\pm$ denote the helicities
of the $W^\pm$)
is only ${O}(\sqrt{s})$.
In the above formulae, $a$ and $b$ are the vector and axial couplings of the
$Z^0$ given by
\be a=\frac{1}{4\sw{}\cw{}}\left( -1+4\sw{2} \right),\qquad
b=-\frac{1}{4\sw{}\cw{}},\label{vacoup}\ee
$\sigma =-\frac{1}{2},\frac{1}{2}$ for lefthanded and righthanded
electrons, respectively, and $\vartheta$
is the center of mass scattering angle.

{}From \eref{mtt} and \eref{mll} we conclude that even with the constraints
\eref{rel} and \eref{couprel} there is
always at least one amplitude which grows as ${O}(s)$,
 unless all three parameters $\dz ,\xg$ and $\yg$ are simultaneously
zero. Thus, the effects of the model we consider lead to a cross
section for $e^+ e^- \rightarrow W^+ W^-$ which grows as $s$
in the terms bilinear (quadratic) in $\dz,\xg$ and $\yg$ and is
${O}(s^0)$ in the linear terms,
while in the standard model the cross section decreases as $s^{-1}$.
The sensitivity to non-standard couplings increases substantially with
increasing energy.


\section{The Term $\bf\O{\bf WB}$\label{termowb}\label{sowb}}
\eqnew\typeout{Section 5}
The interaction term $\O{WB}$ in \eref{owb} contains mixing in the
neutral gauge-boson sector, $W_{\mu\nu}^3 B^{\mu\nu}$, as well as
 non-standard trilinear
interactions, $W_\mu^+ W_\nu^- B^{\mu\nu}$, between the $B$ field and the
charged vector bosons. It also contains non-standard Higgs-interactions.
We will see that the Lagrangian resulting by adding $\O{WB}$
to the standard Lagrangian,
apart from the non-standard Higgs-interactions, coincides with the
Hung-Sakurai model \cite{hs}. In other words, the term $\O{WB}$ essentially
yields an extension of this latter model which is invariant under local
\suu\ transformations quite similar to the terms $\O{W\Phi}$ and
$\O{B\Phi}$ which yield a locally invariant extension of the KMSS
model.

\subsection{$\bf W_{\bf\mu\nu}^{\bf 3}B^{\bf\mu\nu}$ Mixing}
To start with, we concentrate on the $W_{\mu\nu}^3 B^{\mu\nu}$ mixing term
and the corresponding modification of the $W^\pm Z^0$ mass relation and the
fermionic couplings
of the vector-bosons.
In the Lagrangian
\bea\lag{SM+WB}&=&\lag{SM}+\ep{WB}\frac{g^2}{M_W ^2}{\cal L}_{WB},\nn
{\rm where}\quad\O{WB}&=&
-\frac{1}{4}\tr(\Phi^\dagger W_{\mu\nu}\Phi \tau_3)B^{\mu\nu}
,\label{defewb}\eea
we consider that part which describes the vector-boson--fermion
interactions and the
terms quadratic in the gauge-boson fields,
\bea\lag{SM+WB,eff}&=&-\frac{1}{4}W_{\mu\nu}^i
W_i^{\mu\nu}-\frac{1}{4}B_{\mu\nu}
B^{\mu\nu}+M_W^2 W^+ _\mu W^{-\mu}-gW_\mu ^i J_i^\mu -g'B_\mu J_Y
^\mu\nn
 & &+\frac{1}{2}M_W^2(B\enspace W^3)^\mu\pmatrix{\left(\tfrac{g'}{g}\right)
^2&-\tfrac{g'}{g}\cr
-\tfrac{g'}{g}&1}\pmatrix{B\cr W^3}_\mu-\frac{\ep{WB}}{2}W^3
_{\mu\nu}B^{\mu\nu}.\label{bwmix}
\eea
The Lagrangian contains a mass mixing
($W_3^\mu B_\mu$) term and a current-mixing
($W_3^{\mu\nu}B_{\mu\nu}$) term.
In \eref{bwmix}, the basic parameters are $M_W,g,g'$ and $\ep{WB}$.

We will see that an appropriate transition to transformed fields,
$W'^3 _{\mu\nu}$ and $B'_{\mu\nu}$, as well as $\tilde{W}^3 _{\mu\nu}
$ and $\tilde{A}_{\mu\nu}$, allows one to rewrite the Lagrangian
\eref{bwmix} in a form that contains either mass mixing only or current
mixing only, respectively. In the current-mixing form the Lagrangian
will be seen to coincide with the Hung-Sakurai Lagrangian.

One easily verifies by direct substitution that the
transformation
\be\pmatrix{B\cr W_3}=\pmatrix{\frac{1}{\sqrt{1-\ep{WB}^2}}&0\cr
\frac{-\ep{WB}}{\sqrt{1-\ep{WB}^2}}&1}\pmatrix{B'\cr W_3 '}
\label{bw3}\ee
removes the current-mixing term in \eref{bwmix}, modifies the mass
matrix by terms that depend on $\ep{WB}$ and introduces an interaction
of the hypercharge field, $B_\mu^\prime$, with the third component of the weak
isospin current, $J_\mu ^3$, i.e., the neutral current part of \eref{bwmix}
becomes
\bea \lag{N}&=&-\frac{1}{4}{W'}_{\mu\nu}^{3} {W_3'}^{\mu\nu}-\frac{1}{4}
B'_{\mu\nu}{B'}^{\mu\nu}\nn & &+\frac{1}{2}M_W^2\pmatrix{B'&W_3'}^\mu
\pmatrix{\frac{\left(\tfrac{g'}{g}+\ep{WB}\right) ^2}{1-\ep{WB}^2}&
-\frac{\tfrac{g'}{g}+\ep{WB}}{\sqrt{1-\ep{WB}^2}}\cr
-\frac{\tfrac{g'}{g}+\ep{WB}}{\sqrt{1-\ep{WB}^2}}&1}
\pmatrix{B'\cr W'^3}_\mu\nn
 & &-gW_3'^\mu J^3 _\mu -g'B'_\mu \left( -\frac{g}{g'}
\frac{\ep{WB}}{\sqrt{1-\ep{WB}}}J_3 +\frac{1}{\sqrt{1-\ep{WB}^2 } } J_Y
\right) ^\mu .\label{bwprime}\eea

Now consider the Hung-Sakurai model \cite{hs}. In this model, global
SU(2) weak isospin symmetry is broken by electromagnetism. This
breaking is of the strength $\lambda$ and is implemented by a
current-mixing term. There is no mass mixing; instead, the
third component of the weak isospin $W$-triplet has the mass
$M_W$, as required by global SU(2) invariance. The Hung-Sakurai
Lagrangian is thus given by
\bea \lag{H.S.}=-\frac{1}{4}W_{\mu\nu}^i W_i^{\mu\nu}
-\frac{1}{4}\tilde F _{\mu\nu}\tilde{F}^{\mu\nu}
+\frac{1}{2}M_W^2 W_\mu^i W_i^\mu
-gJ_\mu ^i W^\mu _i -eJ^{e.m.}_\mu \tilde{A}^\mu+\lag{mix},\nn
\lag{mix}=-\frac{1}{2}\lambda\tilde{F}_{\mu\nu}\tilde{W}_3^{\mu\nu},
\quad W=(W_1,W_2,\tilde{W}_3),\label{laghs}\eea
where
\be J^{e.m.}=J^3 +J^Y.\ee
To see the equivalence of the Lagrangians in \eref{bwmix}, \eref{bwprime}
with \eref{laghs}, the mass-mixing term in \eref{bwprime} has to be
replaced by a current-mixing term of appropriate strength. The
replacement is achieved by applying a transformation to the fields in
\eref{bwprime} which coincides
with the inverse of \eref{bw3}, apart from replacing
$\epsilon_{WB}$ by a new parameter, $\lambda$, viz.,
\be\pmatrix{B'\cr W_3 '}=\pmatrix{\sqrt{1-\lambda ^2}&0\cr
\lambda&1}\pmatrix{\tilde{A}\cr \tilde{W}_3}.\label{bwprimetrans}\ee
The parameter $\lambda$ has to be chosen such
that the mass mixing disappears and the primordial photon field,
$\tilde{A}$,
couples to the electromagnetic current. Upon substituting
\eref{bwprimetrans} into \eref{bwprime}, one finds that indeed
\eref{bwprime} is converted into the form \eref{laghs}, provided
$\lambda$ is chosen as
\be\lambda=\lambda(g,g',\ep{WB})=\frac{\tfrac{g'}{g}+\ep{WB}}{\sqrt{1+2
\frac{\textstyle g'}{\textstyle g}\ep{WB}+\left(\tfrac{g'}{g}\right)^2
}}\label{lambda}\ee
and $e$ is identified with
\be e=e(g,g',\ep{WB})=g'\tfrac{\sqrt{1-\lambda ^2}}{\sqrt{1-\ep{WB}^2}}
.\label{e}\ee
Passing to $e,g,M_W$ and $\lambda$ as basic parameters of the theory and
using \eref{e} to eliminate $g'$, we may express $\lambda$ as
\be\lambda=\lambda (e,g,\ep{WB})=
\frac{e}{g}(1-\ep{WB}^2)+\ep{WB}\sqrt{1-\left(\frac{e}{g}\right)^2
(1-\ep{WB}^2)} \label{lofep}.\ee
Using this form of $\lambda$ allows one to represent the Hung-Sakurai
Lagrangian \eref{laghs} in terms of $e,g,M_W$ and the
$B^{\mu\nu}W^3_{\mu\nu}$ current-mixing strength $\ep{WB}$ in
\eref{bwmix}.
We note the linear approximation of \eref{lofep},
\bea \lambda
&=&\tfrac{e}{g}+\ep{WB}\sqrt{1-\tfrac{e^2}{g^2}}+O(\ep{WB}^2 ). \eea
On the other hand,
\bea \lambda &=&\tfrac{e}{g}(1-\ep{}),\label{edef}\eea
where $\ep{}$
is identical to the parameter introduced in \cite{su2v}. Thus
\bea \ep{}&=&-\ep{WB}\left( \sqrt{\tfrac{g^2}{e^2}-1+\ep{WB}^2}-\ep{WB}\right)
\nn&=&
-\ep{WB}\tfrac{g}{e}\sqrt{1-\tfrac{e^2}{g^2}}+O(\ep{WB}^2 ).\label{etoewb}\eea
The parameter $-\ep{}$ coincides
\cite{su2v} with the non-standard contribution
to the frequently used parameter $\ep{3}$ \cite{alta}.

The transition to the physical fields, i.e.,
simultaneous diagonalization of the
mass and the kinetic term \cite{bilschi}, may be carried out
either by applying the transformation
\be\pmatrix{B\cr W_3}=\pmatrix{\sqrt{1-2\egl +\left(\tfrac{e}{g}\right)
^2}&-\tfrac{\sqrt{1-2\egl
+\left(\tfrac{e}{g}\right)^2}\:{\textstyle \lambda}}{\sqrt{1-\lambda ^2}}\cr
\tfrac{e}{g}&\tfrac{1-\egl }{\sqrt{1-\lambda ^2}}}
\pmatrix{A\cr Z}\label{simdiag}\ee
to the Lagrangian \eref{bwmix} or by applying
\be\pmatrix{\tilde{A}\cr \tilde{W}^3}=\pmatrix{1&-\tfrac{\lambda}{\sqrt
{1-\lambda ^2}}\cr 0&\tfrac{1}{\sqrt{1-\lambda ^2}} }
\pmatrix{A\cr Z}\label{aztaz}\ee
to the Lagrangian \eref{laghs}.
One obtains\footnote{This effective Lagrangian is the same as the one
obtained (in a certain limit) from
a theory with an additional vector-boson triplet,
$V^\pm ,V^0$, associated with an $SU(2)_V$ local symmetry
\cite{su2v,su2vth}.}
(in the neutral sector)
\bea\lag{N}&=&-\frac{1}{4}Z_{\mu\nu}Z^{\mu\nu}
+\frac{1}{2}\frac{M_W^2}{1-\lambda ^2 }Z_\mu Z^\mu\nn
&&-\frac{1}{4}F_{\mu\nu}F^{\mu\nu}
-\frac{g}{\sqrt{1-\lambda ^2 }}
(J^3-\egl J^{e.m.})^\mu Z_\mu-eJ^{e.m.}_\mu A^\mu.\label{lhs}\eea
Clearly, \eref{lhs} differs from the standard theory by the parameter
$\lambda$, i.e., we have $M_W,e,g$ and $\lambda(e,g,\ep{WB})$
as basic parameters.

{From} \eref{lhs}, one reads off that $\sw{2}$, defined as the negative
ratio of the coefficients of $J^{e.m.}$ and $J^3$ in \eref{lhs},
is in this model given by \be\sw{2}=
\egl,\label{sw2}\ee while the mass of the $Z^0$ turns out to be
\be M_Z^2=\tfrac{M_W^2}{1-\lambda ^2}.
\label{massratio}\ee

Introducing the Fermi coupling, $G_F$,
\be g^2 =\tfrac{8 M_W^2 G_F}{\sqrt{2}},\label{cb}\ee
and using \eref{sw2} and \eref{massratio}, we may replace the input
parameters $e,g,\lambda$ and $M_W$ by the set $\alpha ,G_F , M_Z$ and
$M_W$. One obtains
\be \sw{2}=\sqrt{\tfrac{\pi\alpha}{\sqrt{2}G_F M_Z^2}\left
(\tfrac{M_Z^2}{M_W^2}-1\right) },\label{swtomz}\ee
or, by solving for $M_W^2$,
\be M_W^2=\tfrac{M_Z^2}{1+\tfrac{\sqrt{2}G_F M_Z^2 \sw{2}}{\alpha \pi}}.
\label{mwtosw}\ee
The relation between $\ep{WB}$ and the set $\alpha ,G_F ,M_Z$ and $M_W$ is
contained in \eref{massratio}, \eref{lofep} and \eref{cb}.
By solving for $M_W^2$ one obtains
\be \tfrac{M_W^2}{M_Z^2}=\tfrac{1-\ep{WB}^2-2B\ep{WB}\sqrt{1-\ep{WB}^2}}{2}
+\left( 1-\ep{WB}^2\right) \sqrt{\tfrac{1}{4}-B^2-\tfrac{B\ep{WB}}{\sqrt{
1-\ep{WB}^2}}},\label{mwmz}\ee
where $B$ is defined by
\be B \equiv\sqrt{\tfrac{\pi\alpha }{\sqrt{2}G_F M_Z^2}}.
\label{bdef}\ee
For small $\ep{WB}$ we have
\be \tfrac{M_W^2}{M_Z^2}=\frac{1}{2}+\sqrt{\tfrac{1}{4}-B^2}
-B\left( 1+\tfrac{1}{\sqrt{1-4B^2}}\right) \ep{WB}+{{O}}(\ep{WB}^2).
\label{mwmzlin}\ee
The parameter $\ep{}$ introduced in \eref{edef} may be written as a
function of $\sw{2}$,
\be \ep{}(\sw{2})=1-\tfrac{\sw{2}B^{-2}}{1+\sw{4}B^{-2}},\label{eptosw}\ee
or, alternatively, as a function of $M_W$,
\be\ep{}(M_W)=1-
\sqrt{\tfrac{\sqrt{2}G_F M_W^2}{\alpha \pi}\left( 1-\tfrac{M_W^2}
{M_Z^2}\right) },\label{eptomw}\ee
and \eref{etoewb} becomes
\be \ep{WB}=-\ep{}\tfrac{\sw{}}{\sqrt{1-\sw{2}}} +O(\ep{}^2
).\label{etoewb2}\ee


\subsection{Observability of $\bf{\cal L}_{\bf WB}$}
When deducing $\ep{}$ from the empirical data on the W-mass, the LEP~1
and the low-$q^2$ data, radiative corrections have to be taken into
account. The dependence of the radiative corrections on the (unknown)
value of the top-quark mass, $m_t$, implies a dependence on $m_t$ \cite{kks} of
the value of $\ep{}$ deduced from the data.
A careful two-parameter fit \cite{su2v}
gave the result
\bea \ep{}&=&(0.48\pm 0.61)\cdot 10^{-2}\nn
m_t&=&(114\pm 32)\,{\rm GeV}\label{fit}.\eea
This result was based on a fit to the combined LEP~1 data \cite{lep},
the W-mass measurement \cite{wmass}, and the low-$q^2$ data
\cite{lowq2}. A new evaluation based on the most recent LEP~1 data will
significantly diminish the error in the determination of $\ep{}$. Such a
fit will be presented elsewhere.\par
In the present paper, we will restrict ourselves to a brief discussion
of the dependence of the error in the determination of $\ep{}$ resulting
from the experimental errors in $M_W$ and $\sw{2}$ keeping $m_t$
fixed.\par When deducing $\ep{}$ from a measurement of the W-mass, one
finds from \eref{eptomw} for the error, $\Delta\ep{}$, of $\ep{}$
\be \Delta\ep{}(M_W)=\tfrac{c^2 -s^2}{s^2 c}\tfrac{\Delta M_W}{
 \sqrt{\rho}M_Z}
 +O(\ep{}\Delta\ep{})\;\simeq\; 0.0289\, {\rm GeV}^{-1}\, \Delta M_W,
 \label{deledelmw}\ee
where $s^2=1-c^2$ is the standard weak angle evaluated with the
replacements (e.g. \cite{kks})
\be \alpha(0)\rightarrow\alpha (M_Z^2)\simeq \tfrac{1}{128.9}\label{alpha}
\ee
and
\be M_Z^2\rightarrow\rho M_Z^2\label{rho},\ee
i.e., $s^2$ is obtained from
\be s^2 c^2 =\tfrac{\pi\alpha (M_Z^2)}{\sqrt{2}G_F\rho M_Z^2}\equiv
\tilde{B}^2
\label{scdef}\ee
by solving for $s^2$,
\be s^2=\tfrac{1}{2}-\sqrt{\tfrac{1}{4}-\tilde{B}^2}.\ee
The replacements \eref{alpha} and \eref{rho} take into account radiative
corrections with sufficient accuracy for the present purpose of
estimating the accuracy with which $\ep{}$ can be determined from the
data. A dependence on $m_t$ enters in \eref{deledelmw} via the
$m_t$-dependence of the $\rho$-parameter (e.g., \cite{kks}). The value of
$\Delta\ep{}$, however, changes by less than 4\% if $m_t$ is varied from
100 GeV to 200 GeV. The numerical evaluation of \eref{deledelmw}, using
the present error in the W-mass measurement \cite{wmass},
\be M_W =80.13\pm 0.27\, {\rm GeV},\ee
and the accuracy in $M_W$ of $\Delta M_W =60\, {\rm GeV}$ to be
expected (e.g., \cite{hilu}) from LEP~2, yields, respectively,
\be\Delta\ep{}(M_W)=\left\{\arraycolsep1mm\begin{array}{ccccr}
0.78\cdot 10^{-2}&\quad{\rm for}\quad&\Delta M_W&=&270\, {\rm GeV}\\
0.17\cdot 10^{-2}&\quad{\rm for}\quad&\Delta M_W&=&60\, {\rm GeV}.
\end{array}\right. \label{deledelmwnum}\ee
The corresponding value of $\ep{WB}$, according to \eref{etoewb2}, is
given by
\be\Delta\ep{WB}\simeq -0.55\,\Delta\ep{}.\ee\par
In an analogous manner, from \eref{eptosw}, we obtain for $\Delta\ep{}$
as determined from the measurement of $\sw{2}$,
\be \Delta\ep{}(\sw{2})=-\tfrac{c^2 -s^2}{s^2}\,\Delta\sw{2} +O(\ep{}
 \Delta\ep{})\;\simeq -2.31\,\Delta\sw{2}.\label{deledelsw}\ee
Here, $\sw{2}$ is the weak mixing angle at the ${\rm Z}^0$-mass obtained
from the vector and axial vector couplings of the ${\rm Z}^0$ to
leptons, i.e.,
\be\sw{2}\equiv\bar{s}^2_W(M_Z^2)=\tfrac{1}{4}\left(
1-\tfrac{g_V^l}{g_A^l}\right).\ee
Inserting for $\Delta\bar{s}^2_W$ the error deduced from the charged
lepton forward-backward asymmetries, using the LEP~1 results given in
\cite{lep} and the most recent ones given in \cite{lep93}, as well as
the future accuracy expected by combining LEP~1 and SLC data \cite{rol},
we obtain, respectively,
\be\Delta\ep{}(\sw{2})=\left\{\arraycolsep0.5mm\begin{array}{cccl}
0.72\cdot 10^{-2}&\;{\rm for}\;&\Delta\bar{s}^2_W=0.0031
&({\rm from}\;A_{FB}^l=0.0138\pm 0.0049)\cite{lep},\\
0.28\cdot 10^{-2}&\;{\rm for}\;&\Delta\bar{s}^2_W=0.0012
&({\rm from}\;A_{FB}^l=0.0164\pm 0.0021)\cite{lep93},\\
0.09\cdot 10^{-2}&\;{\rm for}\;&\Delta\bar{s}^2_W=0.0004
&\cite{rol}.
\end{array}\right. \label{deledelswnum}\ee\par
With the assumption that the value of $\ep{}$ obtained from a careful
fit of the available data will allow for $\ep{}=0$ to lie within the
error bars given by \eref{deledelmwnum} and \eref{deledelswnum}, the strength
of a possible ${\rm SU}(2)_{\rm L}$ violation in vector-boson fermion
couplings, $\ep{}\cdot e$ \cite{kks}, will
be restricted in magnitude to a value of the order of $0.2\cdot 10^{-2}e$.


\subsection{Trilinear Gauge-Boson Self-Interactions Induced by
$\bf{\cal L}_{\bf WB}$}
The term $\O{WB}$ has direct and indirect effects on
the self-couplings, leading to non-standard values of $\gz,\kg$ and $\kz$.
The direct effects are due to the
trilinear
interactions $B_{\mu\nu}W^{+\mu}W^{-\nu}$ contained in ${\cal L}_{WB}$.
The indirect effects are due to mixing between $W_3 ^\mu$ and $B^\mu$,
which also affects the self-couplings in $\O{SM}$.
With $\dz$ being defined as
the deviation from the standard case, viz.,
\bea \dz&\equiv&\gz(e,g,\lambda )-\gz(e,g,\lambda =\frac{e}{g})\nn
&=&\gz(e,g,\lambda)-\sqrt{\left(\tfrac{g}{e}\right) ^2 -1},\label{dzredef}\eea
one finds
\bea\dz &=&\frac{g}{e\sqrt{1-\lambda ^2}}\left( 1-\egl
\right)-\sqrt{\left(\frac{g}{e}\right) ^2 -1}, \nn
\xg &=&\gel -1,\nn
\xz &=&-\frac{\lambda}{\sqrt{1-\lambda ^2}}\left( \gel
-1\right),
\label{wbcoup0}\eea
i.e., using \eref{massratio},\ $\xz$ is given in terms of $\xg$ by
\be\xz =-\xg\sqrt{\left( \frac{M_Z}{M_W}\right) ^2 -1}.\label{xzxg}\ee
Such a proportionality of $\xz$ to $\xg$ is expected, as intrinsic ${\rm
SU(2)}_{\rm L}$ violation (of the form \eref{iso}) is absent in
$\O{WB}$.
Also passing to the basic input parameters $\alpha,G_F,M_Z$ and $M_W$ in
$\dz$ and $\xg$, we have
\bea\dz&=&
\frac{1}{\sw{2}}\left[ \cw{2}\sqrt{\left(\frac{M_Z}{M_W}\right) ^2 -1}
-\sqrt{1-\left(\frac{M_W}{M_Z}\right) ^2 -\sw{4}}\,\right]\nn
\xg&=&\tfrac{1}{\sw{2}}\left( 1-\left(\tfrac{M_W}{M_Z}\right) ^2\right) -1,
\label{anom0}\eea
where $\sw{2}$ in terms of $\alpha,G_F,M_Z$ and $M_W$ is given by
\eref{swtomz}. Alternatively, by substituting $M_W$ from \eref{mwtosw},
one may express \eref{xzxg}, \eref{anom0} in terms of $\alpha,G_F,M_Z$
and $\sw{2}$ as basic parameters. Note that all parameters in
\eref{xzxg}, \eref{anom0} can be measured independently of observing
$e^+ e^-\to W^+ W^-$. Obviously, the standard model self-interactions,
$\dz =\xg =\xz =0$,
are recovered if the standard relation
\be \sw{2}=1-\tfrac{M_W^2}{M_Z^2}\ee is fulfilled. In this case, with
\eref{swtomz}, \eref{mwtosw} $M_W$ and $\sw{2}$ are determined by
$\alpha,G_F$ and $M_Z$ alone.

We note that also the vector-boson
self-interactions \eref{xzxg} and \eref{anom0} following from the
Lagrangian \eref{defewb} are the same as those obtained in the Hung--Sakurai
model \cite{hs}.
There, the vector-boson self-interactions are introduced by replacing
the field-strength tensor $W_{\mu\nu}$ in \eref{laghs} by the non-Abelian
one and then coupling the primordial photon field,
$\tilde{A}^\mu$, minimally to the $W^\mu$ field via replacing the
derivatives on the $W$ field by
covariant ones with respect to the electromagnetic gauge freedom.
\par Also with respect to the quadrilinear vector-boson self-interactions,
this minimal substitution procedure of the Hung-Sakurai model leads to
the same results as implied by the Lagrangian \eref{defewb}.


\subsection{Effects of $\bf{\cal L}_{\bf WB}$ on $\bf e^+e^-\to W^+W^-$}
The modifications in the trilinear couplings of the vector bosons among
each other and of the vector bosons to the fermions caused by the
presence of $\O{WB}$ lead to deviations from the standard model
predictions in the high-energy behaviour of the amplitudes for
the process $e^+ e^- \rightarrow W^+ W^-$.
The deviations are fairly modest, however. The high-energy behaviour of
the helicity amplitudes, both for the production ot two longitudinal and
of two transverse W-bosons, is the same as in the standard case, i.e.,
${O}(s^0)$. This decent high-energy behaviour in the presence of the
non-standard couplings \eref{xzxg} and \eref{anom0} is due to
cancellations among the non-standard effects arising from the
fermion--vector-boson and the trilinear bosonic vertices. The only
deviation from the standard-model high-energy behaviour occurs in the
amplitudes for the production of one longitudinal and one transverse
W-boson which grow as $\sqrt{s}$ and, for $s\gg 4\,M_W^2$, are given by
\be {\cal M}_{LT}(\sigma ,\tau ,q)=\tfrac{e^2}{2\sqrt{2}}\tfrac{\sqrt{s}
M_Z^2}{
M_W^2}F(\sigma )\left(q -\sigma\tau\cos\vartheta
\right),\label{mlths}\ee
where
\bea F\left( 1\right)&=&\gel -1=-\ep{}\nn
F\left( -1\right)&=&\frac{3}{2}\gel -1-\frac{1}{2}\left(\gel
\right) ^2=-\tfrac{\ep{}}{2}+{O}(\ep{}^2).\eea
In \eref{mlths}, $\sigma =-1$ and $\sigma =+1$ refer to lefthanded and
righthanded electrons, respectively. The helicity of the transverse
${\rm W}^\pm$ is denoted by $\tau =\pm 1$. Finally, the amplitude also
depends on which one of the two W-bosons, the one with charge $q=+1$ or the
one with charge $q=-1$, is in the transverse polarization state. As the
amplitude in \eref{mlths} is the only one with non-standard high-energy
behaviour, and as the growth with $\sqrt{s}$ is a modest one, a
determination of $\ep{}$ (or, equivalently, $\ep{WB}$,)
from the
non-standard high-energy behaviour
will, at LEP~2 or NLC energies, not reach the above-mentioned
precision based on measurements of $\sw{2}$ and the W-mass.\par
We note that the good high-energy behaviour of the model with $\O{WB}$
also follows from its
equivalence (apart from the Higgs-interactions)
to the Hung-Sakurai model. This latter model is known \cite{hs} to have good
high-energy behaviour apart from the $\sqrt{s}$ growth of the
longitudinal-transverse amplitude in \eref{mlths}. The fairly good
high-energy behaviour, accordingly, cannot be traced back to an
underlying \suu\ gauge invariance (in disagreement with a statement in
\cite{ruj}), as this invariance is violated in the Hung-Sakurai model.
In fact, it is stressed by these latter authors that the absence of
${O}(s)$ terms even holds when, as a consequence of
$\lambda\neq e/g$, \suu\ symmetry is violated.
It is true, however, that any departure from \eref{wbcoup0},
\eref{xzxg}
leads to a considerably stronger growth \cite{ruj} of the
high-energy helicity amplitudes than shown in \eref{mlths}.


\subsection{Gauge-Boson Self-Interactions in the Full Model}
We turn to a brief analysis of the full Lagrangian
\be \lag{eff}=\lag{SM}+(
\ep{W}\frac{g}{M_W^2}\O{W}+\ep{W\Phi}\frac{g}{M_W^2}\O{W\Phi}
+\ep{B\Phi}\frac{g'}{M_W^2}\O{B\Phi}+\ep{WB}\frac{g^2}{M_W^2}\O{WB}).
\label{xyz}
\ee
which includes $W^3 B$ mixing.

\subsubsection{Trilinear Gauge-Boson Self-Interactions}
It will be convenient to use $\lambda (e,g, \ep{WB})$ from \eref{lofep}
instead of $\ep{WB}$ when giving the expressions for the
trilinear couplings. In terms of $e,g,\lambda (e,g, \ep{WB})$ and
$\ep{B\Phi}, \ep{W\Phi}$ and $\ep{W}$, one finds
\bea\dz &=&\frac{g}{e\sqrt{1-\lambda ^2}}\left( 1-\egl +\ep{W\Phi}
\right)-\sqrt{\left(\frac{g}{e}\right) ^2 -1} \nn
\xg &=&\gel -1+\ep{W\Phi}+\ep{B\Phi}\nn
\xz &=&-\frac{\lambda}{\sqrt{1-\lambda ^2}}\left( \gel
-1+\ep{W\Phi}+\ep{B\Phi}\right)\nn
\yg &=&\ep{W}\nn
\yz &=&
\ep{W}\frac{g}{e}\frac{1-\egl}{\sqrt{1-\lambda ^2}}.
\label{wbcoup}\eea
One immediately convinces oneself that the non-standard couplings
previously given in \eref{parameters1}, \eref{rel2} and
\eref{parameters2},
\eref{couprel} are recovered from \eref{wbcoup} in the limit of
$\lambda = e/g = \sw{}$ which corresponds to $\ep{WB} = 0$ (compare
\eref{lofep} and \eref{sw2}). Even for $\lambda \not = e/g\; (\ep{WB}
\not= 0 )$, however, from \eref{wbcoup}, one immediately sees that
$x_Z$ is proportional to $x_\gamma$, while $y_Z$ is proportional to
$y_\gamma$. The proportionality of these couplings, previously (for
$\ep{WB} =0$) given in \eref{rel2} and \eref{couprel}, thus remains
valid even in the presence of the addition of $\O{WB}$ in
\eref{xyz}. Note that this result (which is {\em not} based on any
approximation) is non-trivial insofar as $\O{WB}$ contains $BW_3$
mixing as well as direct non-standard $B\,W^+ W^-$ couplings.
Using \eref{sw2} and \eref{massratio} the proportionality
relations contained in \eref{wbcoup} become
\bea\xz &=&-\xg\sqrt{\left( \frac{M_Z}{M_W}\right) ^2 -1},\nn
\yz &=&\yg\frac{\cw{2}}{\sw{2}}\sqrt{\left(\frac{M_Z}{M_W}\right) ^2 -1}
.\label{couprelgen}\eea
They only differ from the case of $\ep{WB} = 0$ in the values of
the proportionality constants which in the present case depend on
$\alpha, G_F, M_Z$ and a fourth free parameter, e.g., $M_W$ (compare
\eref{mwmz}) or
$\sw{2}$ (compare \eref{swtomz}). As $\alpha, G_F, M_Z$ as well as
$M_W$ can be determined independently of a measurement of the
couplings of the vector bosons to each other, and as $\ep{WB} \not=0$
only changes the magnitude of the factors in \eref{couprelgen}, the
presence or absence of $\O{WB}$ in \eref{xyz} is fairly irrelevant for the
phenomenology of the couplings among the vector bosons.
Quite independently of the presence or absence of $\ep{WB}$, there are
three independent trilinear couplings, $\dz,\xg$ and $\yg$, while $\xz$
and $\yz$ are proportional to $\xg$ and $\yg$, respectively. This
conclusion is quantitavely,
but not qualitatively, changed by the fact that actually $\ep{WB}$ is
strongly constrained anyway (compare \eref{fit}) by LEP~1 measurements.

\subsubsection{Quadrilinear Gauge-Boson Self-Interactions}
The above argument goes through for quadrilinear gauge-boson
self-interactions as well. This is, however, not so surprising,
as the
presence of $\O{WB}$ has only indirect influence on these couplings,
in distinction from the case of the trilinear couplings.
The couplings given by \eref{quartic} can be expressed in terms of the two
free parameters $\ep{W\Phi}$ and $\ep{W}$, and they generalize in the
full model to
\bea\lag{4GB}(\lambda,\ep{W\Phi},
\ep{W})&=&-e^2(A_\mu A^\mu W^+_\nu W^{-\nu} -A_\mu A_\nu W^{+\mu}
W^{-\nu})\nn&&
-2e^2\sqzwe\cotw{2}\left(1+\tfrac{\ep{W\Phi}}{\cw{2}}\right)\nn&&\qquad
(A_\mu Z^\mu W^+_\nu W^{-\nu}-
\frac{1}{2}A_\mu Z_\nu (W^{+\mu} W^{-\nu}+W^{-\mu} W^{+\nu}) )
\nn&&-e^2\zwe\cotw{4}\left(1+2\tfrac{\ep{W\Phi}}{\cw{2}}\right)\nn&&\qquad
(Z_\mu Z^\mu W^+_\nu W^{-\nu} -Z_\mu Z_\nu W^{+\mu} W^{-\nu})\nn&&
+\frac{g^2}{2}\left(1+2\ep{W\Phi}\right)
(W^+_\mu W^-_\nu W^{+\mu}W^{-\nu}-W^+_\mu W^-_\nu W^{-\mu}W^{+\nu})\nn&&
+e^2\frac{\ep{W}}{M_W^2}\bigg\{
[A_\mu A^{\mu\lambda}(W^{+\nu}W^-_{\nu\lambda}+
W^{-\nu}W^+_{\nu\lambda})\nn&&\qquad\qquad
+A_\mu A_{\nu\lambda}
(W^{+\nu}W^{-\lambda\mu}+
W^{-\nu}W^{+\lambda\mu})]\nn&&\qquad
+\sqzwe\cotw{2}
[(A_\mu Z^{\mu\lambda}+Z_\mu A^{\mu\lambda})(W^{+\nu}W^-_{\nu\lambda}+
W^{-\nu}W^+_{\nu\lambda})\nn&&\qquad\qquad
+(A_\mu Z_{\nu\lambda}+Z_\mu A_{\nu\lambda})
(W^{+\nu}W^{-\lambda\mu}+
W^{-\nu}W^{+\lambda\mu})]\nn&&\qquad
+\zwe\cotw{4}
[Z_\mu Z^{\mu\lambda}(W^{+\nu}W^-_{\nu\lambda}+
W^{-\nu}W^+_{\nu\lambda})\nn&&\qquad\qquad
+Z_\mu Z_{\nu\lambda}
(W^{+\nu}W^{-\lambda\mu}+
W^{-\nu}W^{+\lambda\mu})]\bigg\}\nn&&
+g^2\frac{\ep{W}}{M_W^2}(W^+_\mu W^{+\mu\lambda}W^{-\nu}
W^-_{\nu\lambda}-W^+_\mu W^{-\mu\lambda}W^{-\nu}
W^+_{\nu\lambda}).\label{quarticgen}\eea
Again, the modifications due to arbitrary $\lambda$ only become manifest
in the mismatch between $1-\sw{2}$ and the mass ratio.


\section{Conclusion \label{con}}
\eqnew\typeout{Section 6}
A priori, upon excluding C- and P-violation, there exist
three dimension-four and two dimension-six trilinear couplings among
the electroweak vector bosons, and there exists a large number of
quadrilinear couplings.\par As is well-known, via the minimum requirement
that global $SU(2)$ symmetry only be broken by electromagnetism, the
number of {\em dimension-four couplings} is reduced from three to two
by relating the weak and electromagnetic dipole moments of the
$W^\pm$ to each other. At the expense of introducing dimension-six
terms with non-standard Higgs-boson couplings, we have shown that
the above two-parameter dimension-four interactions can be embedded
into a spontaneously broken $SU(2)_L\times U(1)_Y$ theory. This
embedding is irrelevant for the phenomenological exploration of
vector-boson self-couplings by direct measurements of vector-boson
production in $e^+e^-$ annihilation. The inclusion of non-standard
Higgs interactions assures decent one-loop corrections to
LEP~1 observables, however, and, accordingly, serves as an example of how the
above two-parameter model for trilinear and quadrilinear boson
couplings can be made consistent with LEP~1 empirical results.\par The
trilinear {\em quadrupole couplings}, in their well-known single-parameter
form, are invariant under local $SU(2)_L\times U(1)_Y$ transformations
and, accordingly, there is no need for non-standard Higgs terms.

The standard $SU(2)_L \times U(1)_Y$ theory with its celebrated
renormalization properties obviously constitutes the simplest
extension of our present empirical knowledge with respect to the
empirically unknown trilinear and quadrilinear interactions of the
vector bosons with one another. Good arguments may indeed be put
forward for the realization of these standard interactions in nature. The
dimension-six three-parameter alternative of the present paper is to be
considered as an example of how non-standard (trilinear) couplings,
only restricted by a minimum amount of ${\rm SU(2)}$ symmetry, can
nevertheless coexist with LEP~1 precision data. Direct
measurements of these couplings will be indispensable, if one desires
to fully reveal the nature of the electroweak vector bosons.

\section*{Acknowledgement}
It is a pleasure to thank M. Bilenky, W. Hollik, F. M. Renard and H.
Spiesberger for useful discussions.



\end{document}